\documentclass[aps,pra,twocolumn,showpacs,floatfix]{revtex4}
\usepackage{epsfig}
\usepackage{graphicx}
\usepackage{dcolumn}
\usepackage{longtable}
\usepackage{amsthm,amsmath}

\usepackage{xcolor}

\begin{document}

\preprint{\today}

\title{Plasma-screening effects on the electronic structure of multiply charged Al ions using Debye and ionsphere models}

\author{$^1$Madhulita Das\footnote{Email: physics.madhulia@gmail.com}, $^2$B. K. Sahoo\footnote{Email: bijaya@prl.res.in} and $^3$Sourav Pal}

\affiliation{$^1$National Institute of Technology, Rourkela, Odisha - 469008, India}
\affiliation{$^2$Theoretical Physics Division, Physical Research Laboratory, Navrangpura, Ahmedabad - 380009, India}
\affiliation{$^3$Indian Institute of Technology Bombay, Powai, Mumbai - 400076, India}

\begin{abstract}
We analyze atomic structures of plasma embedded aluminum (Al) atom and its ions in the weakly and strongly coupling regimes. The plasma 
screening effects in these atomic systems are accounted for using the Debye and ion sphere (IS) potentials 
for the weakly coupling and strongly coupling plasmas, respectively. Within the Debye model, special attention is given to investigate 
the spherical and non-spherical plasma-screening effects considering in the electron-electron interaction potential. The relativistic 
coupled-cluster (RCC) method has been employed to describe the relativistic and electronic correlation effects in the above atomic 
systems. The variation in the ionization potentials (IPs) and excitation energies (EEs) of the plasma embedded Al ions are presented. 
It is found that the atomic systems exhibit more stability when the exact screening effects are taken into account. It is also showed 
that in the presence of strongly coupled plasma environment, the highly ionized Al ions show blue and red shifts in the spectral lines 
of the transitions between the states with same and different principal quantum numbers, respectively. Comparison among the results 
obtained from the Debye and IS models are also carried out considering similar plasma conditions.
\end{abstract}


\maketitle

\section{Introduction}

Electronic structures of atomic systems immersed in hot and dense plasma environment may be remarkably different from their corresponding 
isolated candidates. Accurate estimation of the electronic structures of atoms or ions is one of the active fields of research in the recent years 
for their wide range of applications \cite{Murillo,Salzman,Lindl}. The plasma may contain different charged species as well as free electrons. These 
charged particles can screen the atomic potential of the embedded atomic systems resulting deviations in the structures of the systems from their
corresponding isolated systems. First principle calculations of electronic structures of the plasma embedded atomic systems are extremely 
strenuous due to complexity involved in the description of the Coulomb potentials. Thus, to describe their structures conveniently, model atomic 
potentials are used which account for the plasma screening effects. The plasma environment can be classified into weakly and strongly coupled plasma depending on the strength of its coupling constant $\Gamma$, which is the ratio of the Coulomb potential energy to the thermal energy. For weakly 
coupled plasma ($\Gamma$ $\ll$ 1; i.e. low density and high temperature), the screening 
effect can be appropriately described using the Debye model \cite{Murillo}. However, in the strongly coupled plasma ($\Gamma$ $\gg$ 1 implying high 
density and low temperature) ion-sphere (IS) potential model \cite{Ichimaru} is the best suited model for accounting the plasma screening 
effects. Both the Debye and ion-sphere models have been successfully employed several times earlier to describe the electronic properties of 
different plasma embedded atomic systems \cite{Qi2008,Saha2005,Qi2009,Saha2002,paul,Saha2009,Das}. The effect of nuclear charge screening by plasma free electrons is reciprocated in form of ionization potential depression (IPD) or continuum lowering \cite{Inglis,Feynman,Ecker,Stewart}. Other crucial spectral properties of atomic systems that are of immense interest are excitation 
energies (EEs), spectral line shifts, line broadenings, energy level crossings etc. Accurate knowledge of these quantities is essential in describing 
equation of state of plasma \cite{Wang}, finding out opacity of an element in the astrophysical plasma \cite{Bailey,Seaton}, in understanding dynamics 
of atomic systems in the laser cooling and trapping of ions \cite{Eschner}, inertial confinement fusion studies \cite{Storm} etc. 

These days many laboratory experiments are being performed in this regard. Recently, aluminum (Al I) and its different multi-charged ions have been 
considered for the laboratory plasma experiments. In the Linac Coherent Light Source (LCLS) and  Free-Electron-Laser (FEL) experiments, Al is used as 
a common target material. John {\it et al} \cite{John} had characterized the absorption spectroscopy of cold and dense Al plasma using a pulsed 
soft X-ray continuum back-lighting source. In their experiment, they characterized the L-shell spectra of Al IV and Al V at the plasma temperature and 
electron density of 12 eV and 0.6$\times$ 10$^{21}$ cm$^{-3}$, respectively. Similarly, Tijerina {\it et al} \cite{Tijerina} had used a wide 
field spectrograph to analyze the behavior of the laser-produced Al-plasma by 
measuring the line widths of the singly (Al II) and doubly (Al III) ionized Al. Ciobanu {\it et al} \cite{Ciobanu} had also studied the spectroscopy 
of Al plasma  by using the second (532 nm) harmonic of a Q-switched pulsed Nd-YAG laser and had observed many line intensities. In other works, Hoarty 
{\it et al.} \cite{Hoarty} and Ciricosta {\it et al.} \cite{Ciricosta} experimentally studied the influence of hot and dense plasma environment on the 
spectroscopy of Al atom and had reported its IPDs. Ciricosta {\it et al} had used X-ray free electron lasers to analyze the K-edge spectra of 
solid-density Al plasma with temperature up to 180 eV. In a different experiment, the newly commissioned Orion laser system has been used to investigate 
the Al samples with a varying plasma density of 1 to 10g/cc and electron temperature of 500 eV and 700 eV to study the IPD as a function of plasma density 
\cite{Hoarty}. Also, a number of theoretical studies on Al plasma have been carried out because of its experimental interest. Zeng {\it et al} 
\cite{Zeng} had performed extensive calculations of the X-ray transmission spectra for the high-power laser-produced Al plasma in local 
thermodynamic equilibrium (LTE) by employing a configuration interaction (CI) scheme and the R-matrix method with the detailed-term-accounting approximation. Feng {\it et al} \cite{Feng} had theoretically simulated the X-ray emission spectra of Li-like Al ion by using the 
collisional radiative model. Preston {\it et al} \cite{Preston} had simulated the emission spectra of the hot and dense Al plasma using the 
Stewarte-Pyatt (SP) and modified EckereKr{\"o}ll (mEK) models. Very recently, Shuji {\it et al} \cite{Shuji} have calculated the radiative opacity of 
the Al plasma in LTE by using the time-dependent density-functional theory (TDDFT) at temperature $T = 20$ eV and density $0.01 ~g/cm^3$. The strongly 
coupled plasma is mostly seen in the highly evolved stars, interior of the Jovian planets, explosive shock tubes, laser produced plasma and inertial 
confinement fusion plasmas \cite{Ichimaru,Nantel,Nazir,Hammel}. Most of these studies have been carried out in the Debye model formalism. However, 
there has been no theoretical investigation of the strongly coupled Al-plasma carried out in the IS model framework. Nevertheless, it appears from a large 
number of studies that probing structures of Al-plasma are of ample interest to both the experimentalists and theoreticians working in this field.

The primary interest of the present work is to carry out an {\it ab initio} investigation of electronic structures of Al and some of its ions in both 
weakly and strongly coupled plasma environment. In most of the previous studies the electronic structures of Al plasma have been investigated using 
many-body methods which account for the electron correlation effects inefficiently. Accurate calculations in the atomic systems with more than four 
electrons require a many-body method that is capable of including electron correlation effects rigorously. Again, the relativistic effects in these 
ions are usually large. In this work, we employ a relativistic coupled-cluster (RCC) method to carry out the theoretical investigations in the 
considered atomic systems. The RCC method is an all order perturbative method that obeys the size extensivity and size consistent behavior 
\cite{szabo,bartlett}. In weakly coupled plasma, we consider the Debye-screened potential instead of the usual atomic potential in the RCC method to 
describe the change in the spectroscopic properties of the plasma embedded Al ions. In the approach, in which, the screening effect is taken into account 
only through the nuclear potential is referred to as the spherical Debye (SD) potential approximation. However, plasma free electrons may also play an
important role 
in the screening of bound electron-electron interaction term in the potential. The approach in which both nuclear and electronic charge screenings are 
considered explicitly is denoted as non-spherical Debye (NSD) potential approximation.
 Owing to the complexity in the consideration of the NSD potential approximation in a perturbative approach, this is rarely applied in the investigation 
 of electronic structure of the plasma embedded atomic systems. The SD potential approximation may  provide reasonably accurate results in the H-like, He-like and 
 Li-like atomic systems, where there are not many electrons present. However, it has been found that consideration of NSD potential approximation can lead 
 to very interesting results in the evaluations of the orbital energies and transition probabilities in the plasma embedded atomic systems \cite{Ondrej,Jung}. 
 Recently, Gutierrez {\it et al} \cite{Gutierrez} had also shown that NSD potential gives rise large collision strengths compared to the SD potential approximation. 
 In this work, we intend to make a comparative analysis of results considering both the SD and NSD potentials through our RCC method. Similarly, the effect 
 of the strongly coupled plasma environment on the atomic structure of Al ions 
is being investigated by considering the IS potential in the RCC method. Again, we consider few cases with the given experimental conditions of plasma and investigate 
IPDs and EEs of the Al III and Al XI ions in both the NSD potential approximation of the Debye model and the IS model to make a comparative analysis of the 
results obtained from these models.

The paper is organized as follows: In Sec. \ref{sec2}, we introduce the screening models that are considered in the calculations for 
the description of the atomic spectra and Sec. \ref{sec3} describes the RCC method briefly. In Sec. \ref{sec4}, we present our results and 
compare with other available data. These results are finally summarized in Sec. \ref{sec5}. Unless stated otherwise, we have used atomic units (a.u.) 
through out the paper.

\section{Plasma Models}\label{sec2}

We describe below the salient features of the models those have been adopted to account the screening effects in the considered Al systems. 
We also give explicitly the expression for the two-body screening Debye potential in the multipole expansion form. Formulas to estimate the 
Debye length for the Debye potential and radius for the IS model are also given.

\begin{table*}[t]
\caption{Calculated ionization potentials (IPs) and excitation energies (EEs) of Al I, Al III, Al IX and Al XI using the CCSD method. These 
values are compared with the available values in the NIST database \cite{NIST}. All the quantities are given in cm$^{-1}$.}\label{tab1}
\begin{ruledtabular}
\begin{tabular}{c c c c c c c c c c c c c c c }
\multicolumn{3}{c}{  Al I } &  &\multicolumn{3}{c}{ Al III } &  &\multicolumn{3}{c}{ Al IX } &  &\multicolumn{3}{c}{ Al XI } \\
\cline{1-3}\cline{5-7}\cline{9-11}\cline{13-15} \\ 
State   & Present  &  NIST  & & State   &   Present & NIST & & State & Present  & NIST   && State & Present & NIST \\
\hline
          &        &                  &&          &           &                & &       &          &                  &&       &         & \\
          \multicolumn{15}{c}{\underline{Ionization Potentials} } \\
          &        &                  &&          &           &                & &       &          &                  &&       &         & \\
3P$_{1/2}$ &47777.96 & 48278.48 && 3S$_{1/2}$      & 229311.73 & 229445.7 &&    2P$_{1/2}$     & 2664482.59 & 2663340   &&  2S$_{1/2}$       & 3565617.36 & 3565010 \\
          &        &                  &&          &           &                & &       &          &                  &&       &         & \\
          \multicolumn{15}{c}{\underline{ Excitation Energies }} \\ 
             &        &                  &&          &           &                & &       &          &                  &&       &         & \\       
3P$_{3/2}$&212.72   & 112.06  && 3P$_{1/2}$ & 53673.29 &  53682.93 && 2P$_{3/2}$ & 5853.17    & 4890   && 2P$_{1/2}$ & 176124.76  & 176019  \\
3D$_{3/2}$&33038.43 & 32435.45&& 3P$_{3/2}$ & 53920.25 & 53916.60  && 3S$_{1/2}$ & 1500857.39 & 1501020 && 2P$_{3/2}$ & 182372.76  & 181808 \\
3D$_{5/2}$&33027.14 & 32436.79&& 3D$_{3/2}$ & 115955.41& 115958.50 && 3P$_{1/2}$ & 1574228.67 &  && 3S$_{1/2}$ & 2020890.43 & 2020450 \\
4S$_{1/2}$&24943.09 & 25347.75&& 3D$_{5/2}$ & 115954.15& 115956.21 && 3P$_{3/2}$ & 1575551.79 &           && 3S$_{1/2}$ & 2069289.06 & 2068770 \\
4P$_{1/2}$&32522.94 & 32949.80&& 4S$_{1/2}$ & 126062.54& 126164.05 && 3D$_{3/2}$ & 1643032.22 & 1642140   && 3P$_{3/2}$ & 2071130.56 & 2070520 \\
4P$_{3/2}$&32543.51 & 32965.64&& 4P$_{1/2}$ & 143537.40& 143633.38 && 3D$_{5/2}$ & 1643379.69 & 1642380   && 3D$_{3/2}$ & 2088662.63 & 2088100 \\
         &         &         && 4P$_{3/2}$ & 143622.20& 143713.50 && 4S$_{1/2}$ & 2043518.52 &  && 3D$_{5/2}$ & 2089195.23 & 2088530 \\
         &         &         &&           &          &           && 4P$_{1/2}$ & 2071861.95 &  && 4S$_{1/2}$ & 2706972.19 & 2705700 \\
         &         &         &&           &          &           && 4P$_{3/2}$ & 2072401.64 &           && 4P$_{1/2}$ & 2726831.30 & 2726120 \\
         &         &         &&           &          &           &&           &            &           && 4P$_{3/2}$ & 2727605.74 & 2726910 \\
\end{tabular}
\end{ruledtabular}
\end{table*}

\subsection{Debye Model}

In the weakly coupled plasma, screened potential experienced by an electron located at $r_i$ in an atomic system due to the presence of other free 
electrons inside the plasma is given as \cite{debye} follows
\begin{equation}
V_{\mathrm{eff}}(r_i)=  e^{-r_{i}/D} V_{nuc}(r_i) + \sum_{j \ge i}^{N}\frac{e ^{-r_{ij}/D}}{r_{ij}},
\label{eqn1}
\end{equation}
where $V_{nuc}(r_i)$ is the nuclear potential, $N$ is the number of bound electrons and $D$ is the Debye screening length. The expression to determine 
$D$ for a plasma having temperature $T$ and ion density $n_i$ is given by  
\begin{eqnarray}
D= \left [ \frac{k_{\rm{B}} T}{4\pi(1+Z)n_i} \right ]^{1/2}
\label{debye}
\end{eqnarray} 
for the Boltzmann constant $k_{\rm{B}}$ and the nuclear charge $Z$.
\begin{figure*}[hbt]
\centering
  \begin{tabular}{@{}c@{}c@{}}
    \includegraphics[width=.5\textwidth]{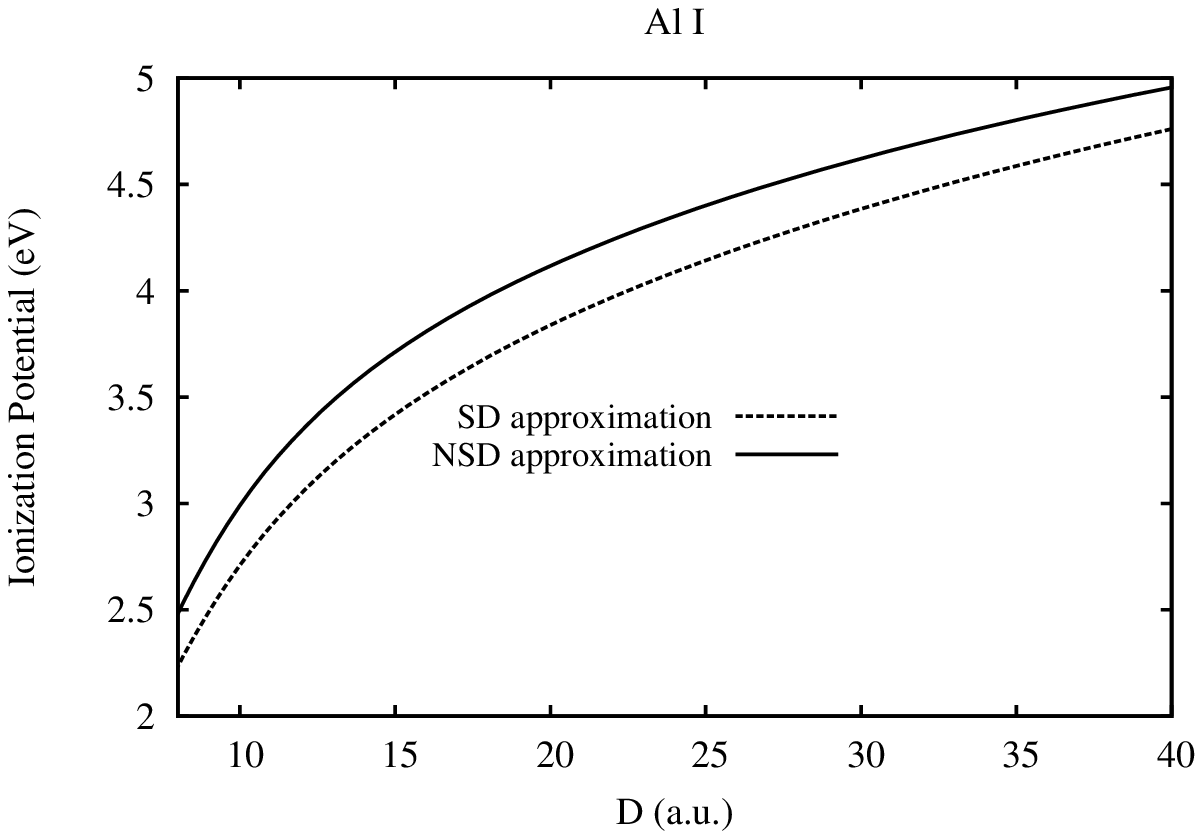}&
    \includegraphics[width=.5\textwidth]{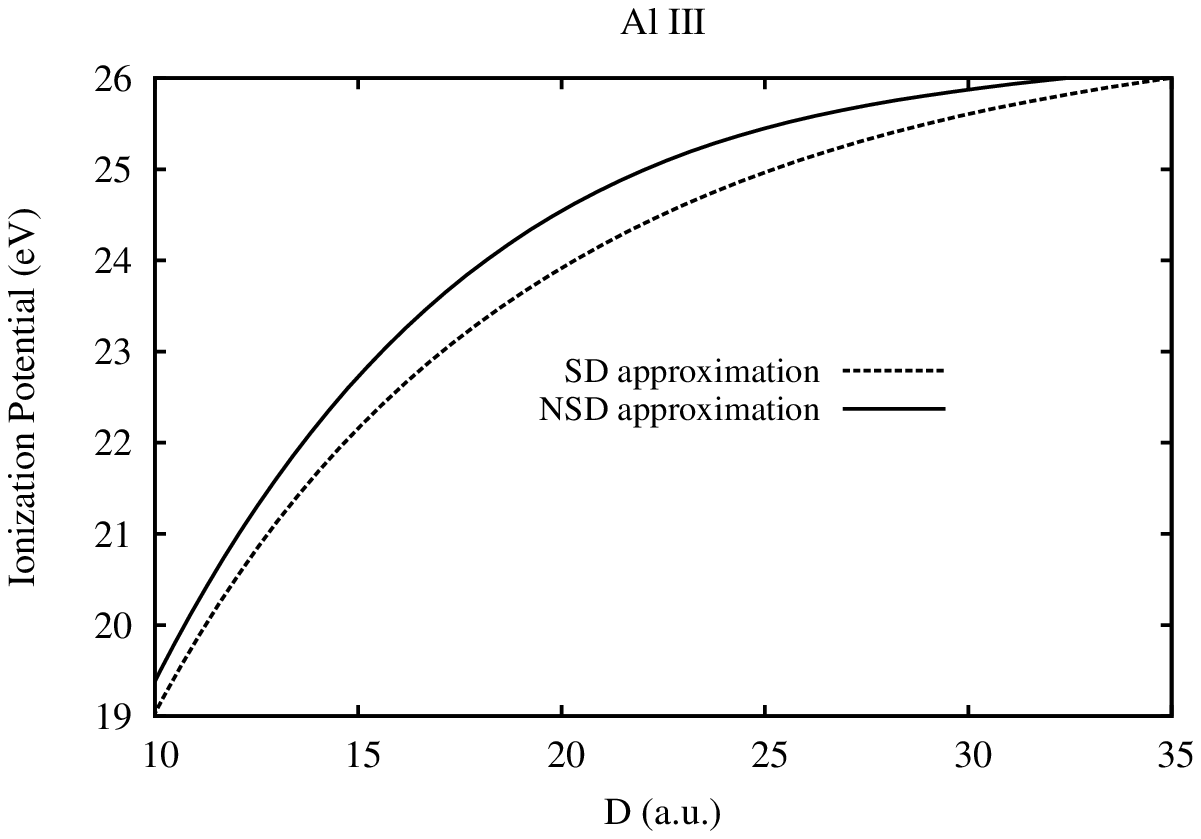}\\
    (a) & (b) \\
    \includegraphics[width=.5\textwidth]{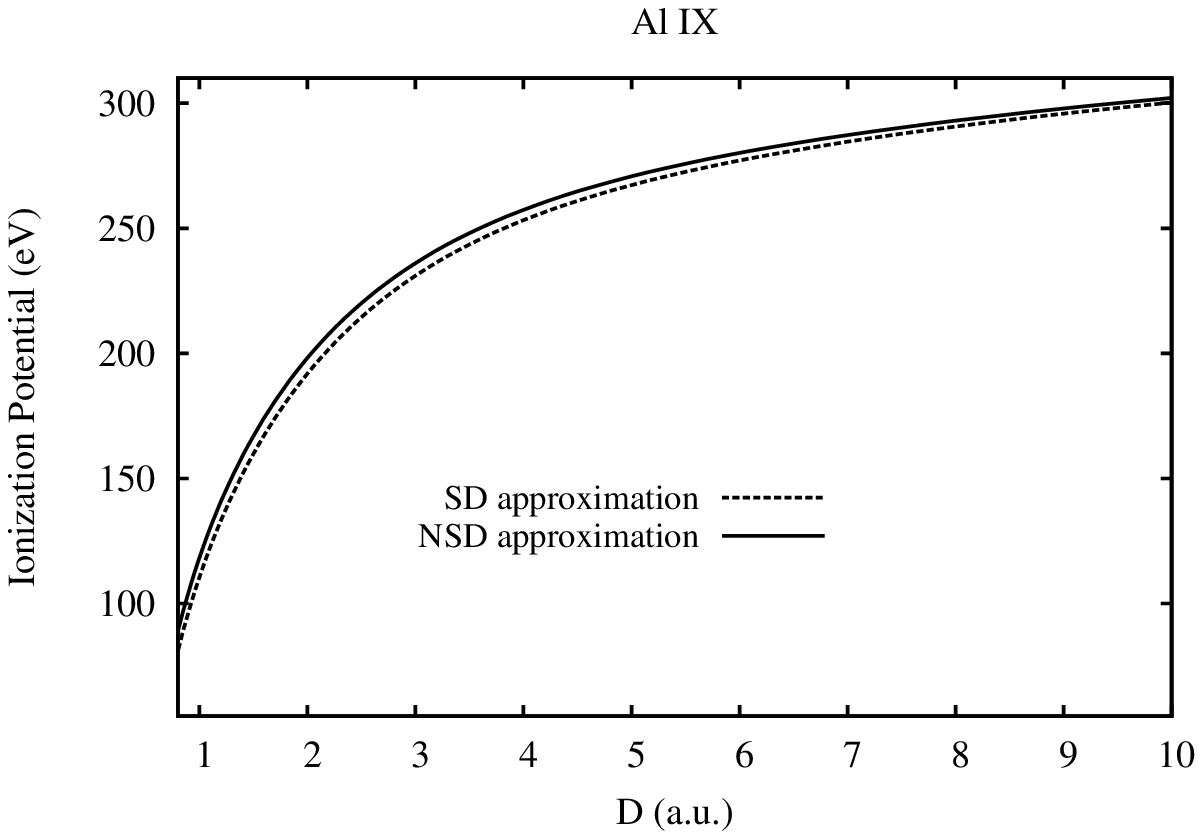}&
    \includegraphics[width=.5\textwidth]{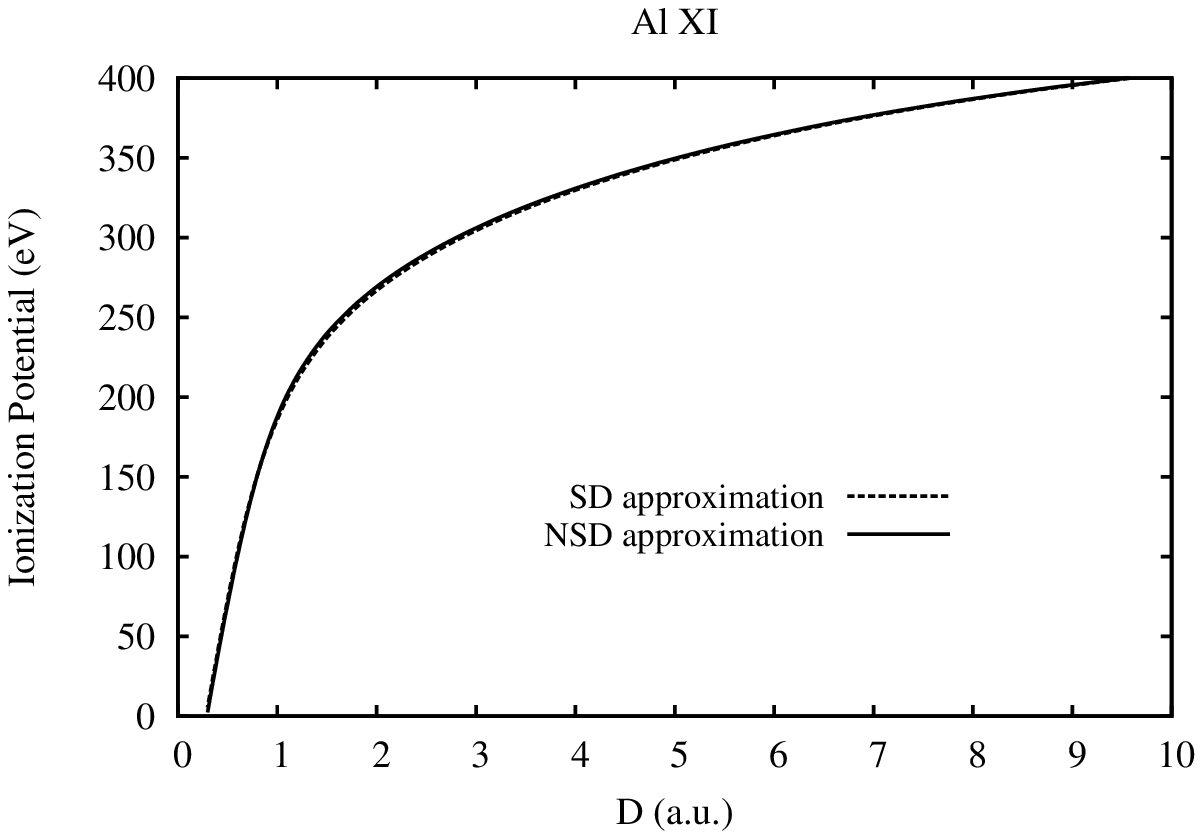}\\ 
    (c) & (d) \\
     \end{tabular}
\caption{Variation in the ionization potentials (IPs) with the Debye screening length in (a) Al I, (b) Al III, (c) Al XI and (d) Al XI.}
 \label{fig1}
\end{figure*}

The Debye potential is a long-range potential where vanishing boundary conditions are satisfied at infinity. Owing to the complicated derivation 
of the two-body screening potential and difficulties to perform their calculations, most of the earlier works, that 
were dealt with the lighter atomic systems, had incorporated the Debye screenings only in the electron-nucleus potential. In our calculations, 
we refer this to SD potential approximation. In this approach Eq. (\ref{eqn1}) is given by
\begin{equation}
V_{\mathrm{eff}}(r_i)= e^{-r_{i}/D} V_{nuc}(r_i) + \sum_{j \ge i}^{N}\frac{1}{r_{ij}}.
\label{eqn2}
\end{equation}
Since the considered Al ions have more than four electrons, it is anticipated that the two-body correlation effects can be significant. So
it is imperative to account for the screening effects in the two-body interaction term. When the exact effective 
potential given by Eq. (\ref{eqn1}) is taken then we refer to the approach as NSD potential approximation. For the comprehensive understanding, we 
consider both the cases to make a comparative study.

Again the nuclear potential $V_{nuc}(r)$ is often estimated for the spectroscopy study of the plasma embedded atomic systems by considering the nucleus as a point-like object. In this case, it yields
\begin{eqnarray}
 V_{nuc}(r_i) = - \frac{Z}{r_i} .
\end{eqnarray}
To have a more realistic potential, we use the standard Fermi-charge distribution to describe the finite size of the nucleus. In this case, we
have \cite{Estevez}
\begin{flushleft}
\begin{eqnarray}
 V_{nuc}(r_i) = -\frac{Z}{\mathcal{N}r_i} \times \ \ \ \ \ \ \ \ \ \ \ \ \ \ \ \ \ \ \ \ \ \ \ \ \ \ \ \ \ \ \ \ \ \ \ \ \ \ \ \  \nonumber\\
\left\{\begin{array}{rl}
\frac{1}{b}(\frac{3}{2}+\frac{a^2\pi^2}{2b^2}-\frac{r^2}{2b^2}+\frac{3a^2}{b^2}P_2^+\frac{6a^3}{b^2r}(S_3-P_3^+)) & \mbox{for $r_i \leq b$}\\
\frac{1}{r_i}(1+\frac{a62\pi^2}{b^2}-\frac{3a^2r}{b^3}P_2^-+\frac{6a^3}{b63}(S_3-P_3^-))                           & \mbox{for $r_i >b$} ,
\end{array}\right.   
\label{eq12}
\end{eqnarray}
\end{flushleft}
where the factors are 
\begin{eqnarray}
\mathcal{N} &=& 1+ \frac{a^2\pi^2}{b^2} + \frac{6a^3}{b^3}S_3  \nonumber \\
\text{with} \ \ \ \ S_k &=& \sum_{l=1}^{\infty} \frac{(-1)^{l-1}}{l^k}e^{-lb/a} \ \ \  \nonumber \\
\text{and} \ \ \ \ P_k^{\pm} &=& \sum_{l=1}^{\infty} \frac{(-1)^{l-1}}{l^k}e^{\pm l(r-b)/a} . 
\end{eqnarray}
Here, the parameter $b$ is known as the half-charge radius and $a$ is related to the skin thickness of the nucleus. These
two parameters are evaluated by
\begin{eqnarray}
a&=& 2.3/4(ln3) \nonumber \\
\text{and} \ \ \ \ \ b&=& \sqrt{\frac {5}{3} r_{rms}^2 - \frac {7}{3} a^2 \pi^2}
\end{eqnarray}
with the appropriate values of the root mean square radius of the nucleus $r_{rms}$, which is estimated using the empirical formula
\begin{eqnarray}
 r_{rms} =0.836 A^{1/3} + 0.570
\end{eqnarray}
in fm for the atomic mass $A$.

The two-body screened potential is expressed as
\begin{eqnarray}
V_{ee}(r_i,r_j) &=& \sum_{j \ge i}^{N}\frac{1}{r_{ij}} e^{-r_{ij}/D} \nonumber \\
       &=& \frac{4\pi}{\sqrt{r_ir_j}}\sum_{l=0}^{\infty}I_{l+\frac{1}{2}}\Big{(}\frac{r_<}{D}\Big{)} K_{l+\frac{1}{2}}\Big{(}
       \frac{r_>}{D}\Big{)} \nonumber \\ && \times \sum_{m=-l}^l Y_{lm}^{\ast}(\theta,\phi)Y_{lm}(\theta,\phi) ,
\label{veff}
\end{eqnarray}
where $I_{l+\frac{1}{2}}$ and $K_{l+\frac{1}{2}}$ are the modified Bessel functions of the first and second kind, respectively. $r_> =$ 
max($r_i,r_j$); $r_<=$ min($r_i,r_j$), and $Y_{lm}(\theta,\phi)$ is the spherical harmonics of rank $l$. The above potential is solved 
in the similar way as the Coulomb potential $1/r_{ij}$ is being evaluated in a common atomic structure calculations. 

\begin{table}
\caption{A list of IPs (in cm$^{-1}$) for some arbitrary values of Debye length ($D$ in a.u.) obtained using the SD and NSD potential 
approximations in the Debye model. Differences in the results from both the approximations are given as $\Delta$IP in cm$^{-1}$.}\label{tab2}
\begin{ruledtabular}
\begin{tabular}{l l r r r r }
Ion & $D$   &  SD  & NSD & $\Delta$IP   \\
\hline
 & & \\
Al I &   75     &  44117.72 & 44875.06  & 757.34          \\
     &   65     &  43571.82 & 44433.20  & 861.38          \\
     &   30     &  38975.59 & 40622.00  & 1646.41         \\
     &   20     &  34993.49 & 37176.15  & 2182.66         \\
     &   10     &  24592.35 & 27444.05  & 2851.70         \\
\hline
& & \\
Al III & 60     & 218698.15 & 218518.13 &180.02       \\
       & 40     & 213538.94 & 213254.15 &284.79        \\
       & 30     & 208475.70 & 208076.78 &398.92         \\
       & 10     & 168520.85 & 166833.28 &1687.57       \\
       & 5      & 122568.03 & 118492.86 &4075.17       \\
       & 3      & 76603.26  & 68983.50  &7619.76         \\
\hline
 & & \\
Al IX  & 10     & 2458618.42 & 2471216.96 &  12598.54    \\
       & 5      & 2262628.87 & 2286219.55 &  23590.68   \\
       & 2      & 1726164.27 & 1773863.71 &  47699.44   \\
       & 1.5    & 1462681.28 & 1519187.01 &  56505.73   \\
       & 1.0    & 1003850.78 & 1071138.01 &  67287.23    \\
       & 0.8    & 710695.14  & 779068.71  &  68373.57    \\
\hline
& & \\
Al XI & 12      & 3368721.32 & 3368613.91 & 107.41\\
      & 5       & 3106688.38 & 3106348.89 & 339.49\\
      & 3       & 2825588.51 & 2824867.62 & 720.89 \\
      & 1.5     & 2201132.03 & 2198937.43 & 2194.6 \\
      & 0.6     & 884330.29  & 874154.68  & 10175.61 \\
      & 0.5     & 598991.96  & 585398.94  & 13593.02  \\
\end{tabular} 
\end{ruledtabular}
\end{table}

\subsection{Ion Sphere Model}

In the strongly coupled plasma, the effective potential of the plasma embedded atomic system is given by \cite{Balazs}
\begin{eqnarray}
 V_{\mathrm{eff}}^{\mathrm{IS}}(r_i) &=& \frac{(\mbox Z-\mbox N)}{2R} \Big [3-\left(\frac {r_{i}}{\mbox R} \right)^2 \Big ] ,
\end{eqnarray}
where $Z$, $N$ and $R$ represent for the nuclear charge, the charge state of the ion and the ion sphere radius, respectively. Here
$R$ is related to the ion density n$_{ion}$ as
\begin{eqnarray}
\mbox R &=& \left(\frac{3}{4 \pi n_{ion}}\right)^{1/3} .
\end{eqnarray}
Note that though it appears in the above expressions as if the plasma temperature dependency is absence, but effects due to plasma temperature 
are taken into account in determining the free electron distribution while deriving the expression for the above radius \cite{Salzman}. Unlike the 
case of Debye model, a finite boundary condition [$\psi(r)|_{\tiny\mbox R}= 0 $] is imposed in the IS model \cite{Balazs}. This boundary condition 
indirectly brings in the effect of the external plasma confinement due to the neighboring ions. 

\begin{figure*}[hbt]
\centering
  \begin{tabular}{@{}c@{}c@{}}
    \includegraphics[width=.5\textwidth]{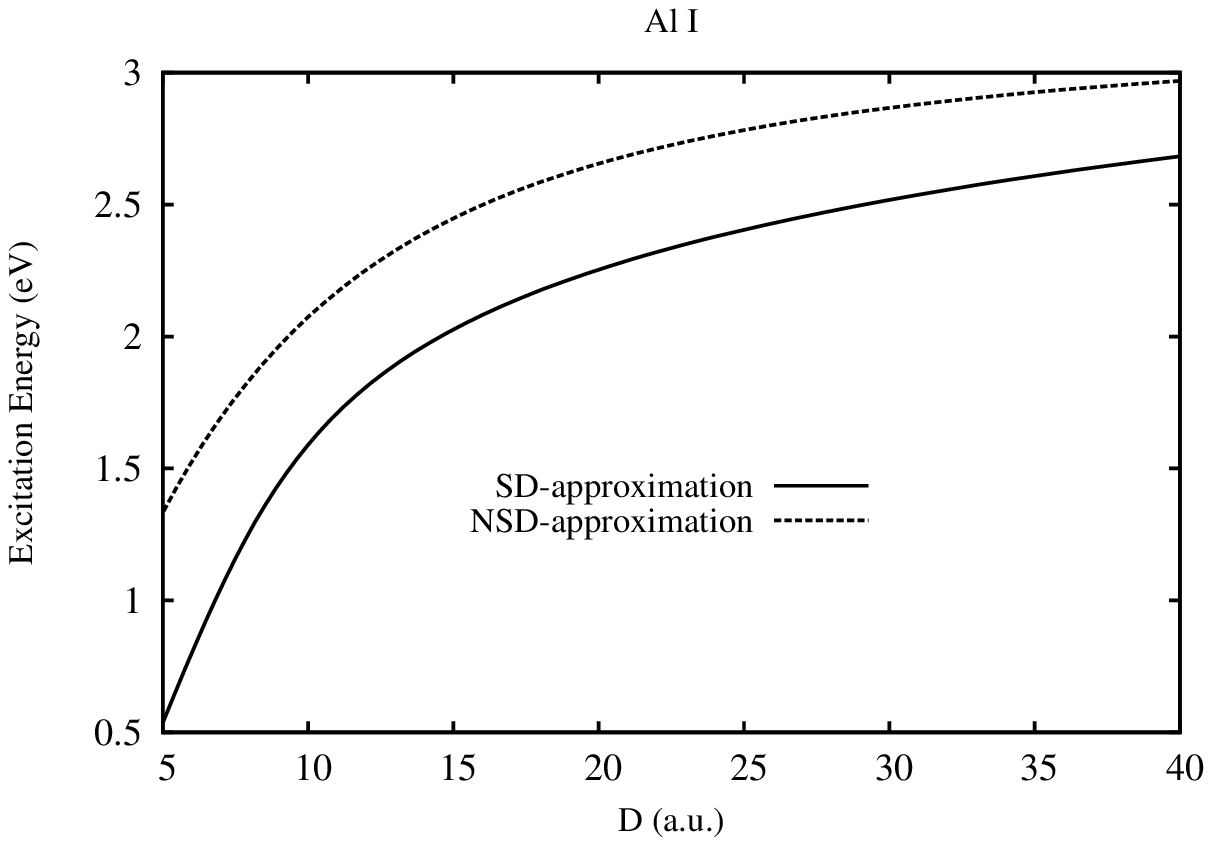}&
    \includegraphics[width=.5\textwidth]{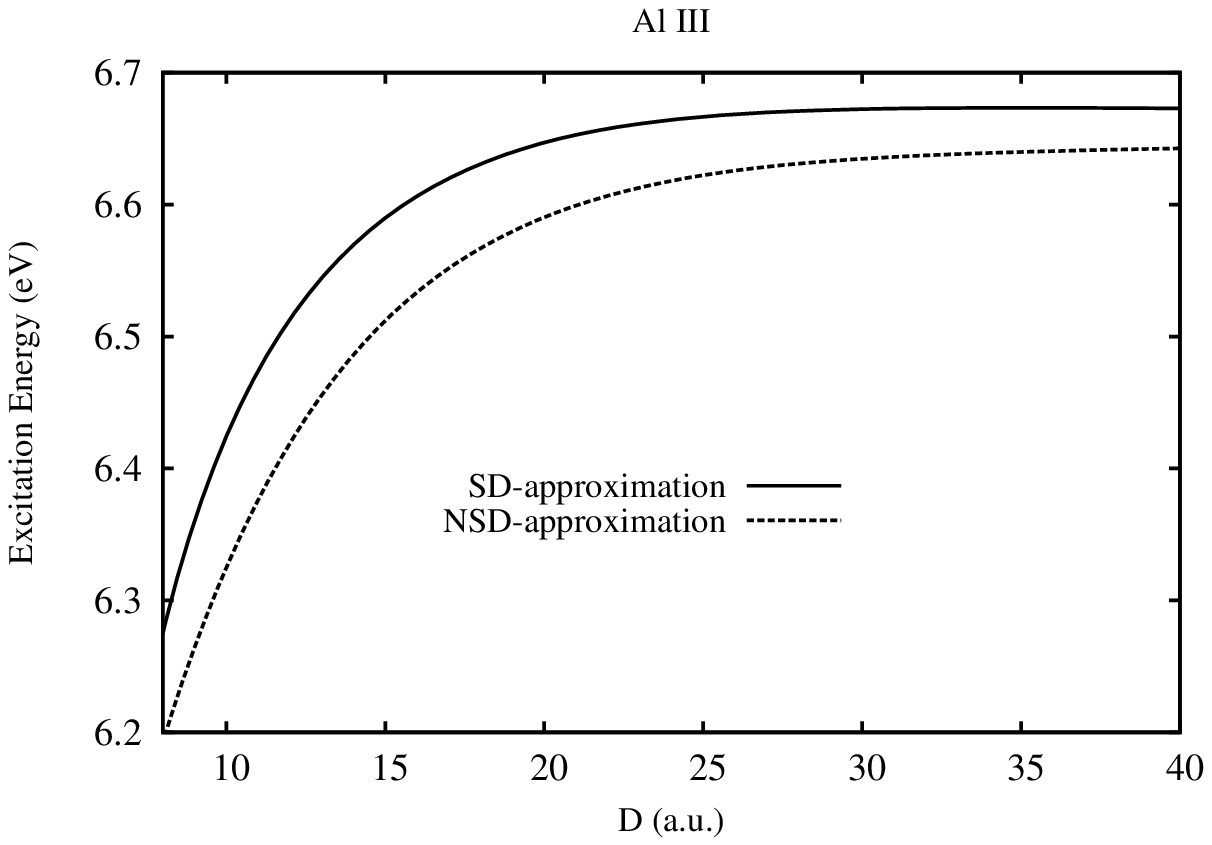}\\
     (i) & (ii) \\
    \includegraphics[width=.5\textwidth]{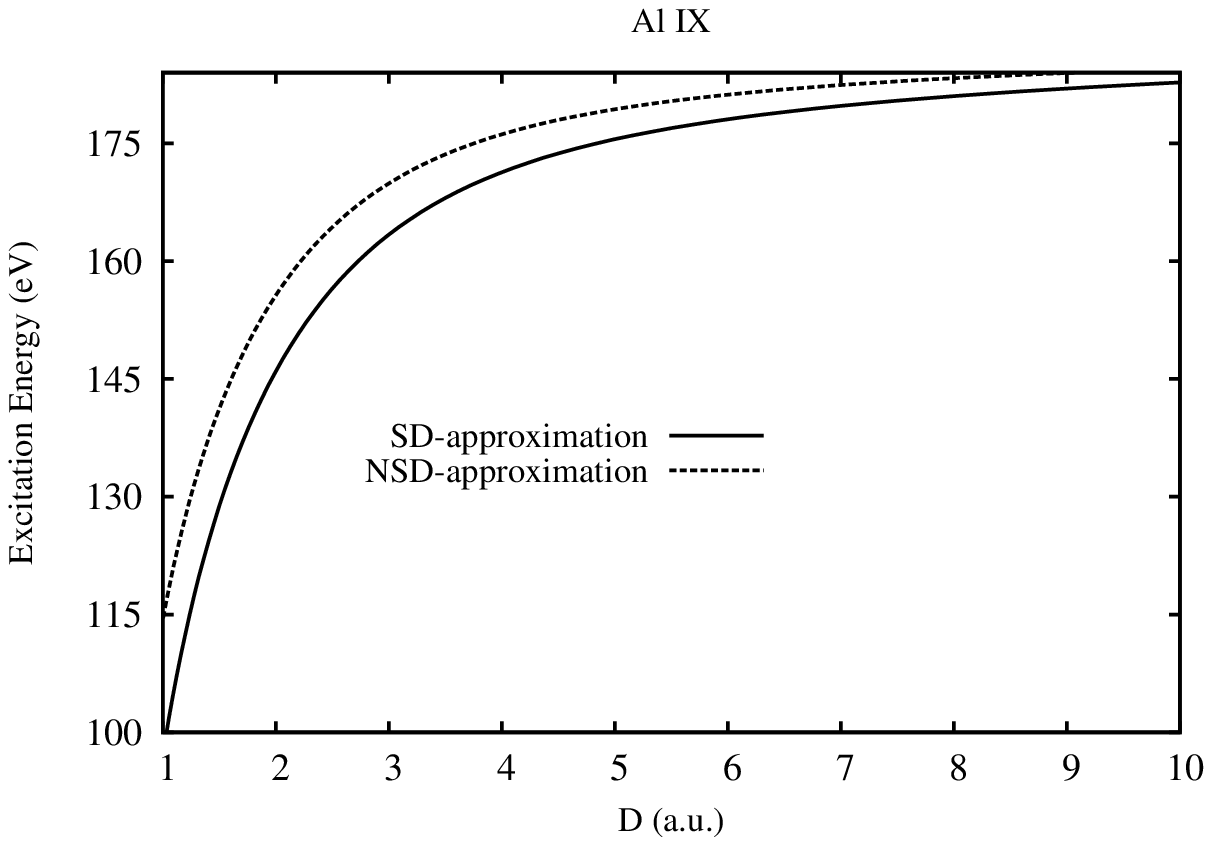}&
    \includegraphics[width=.5\textwidth]{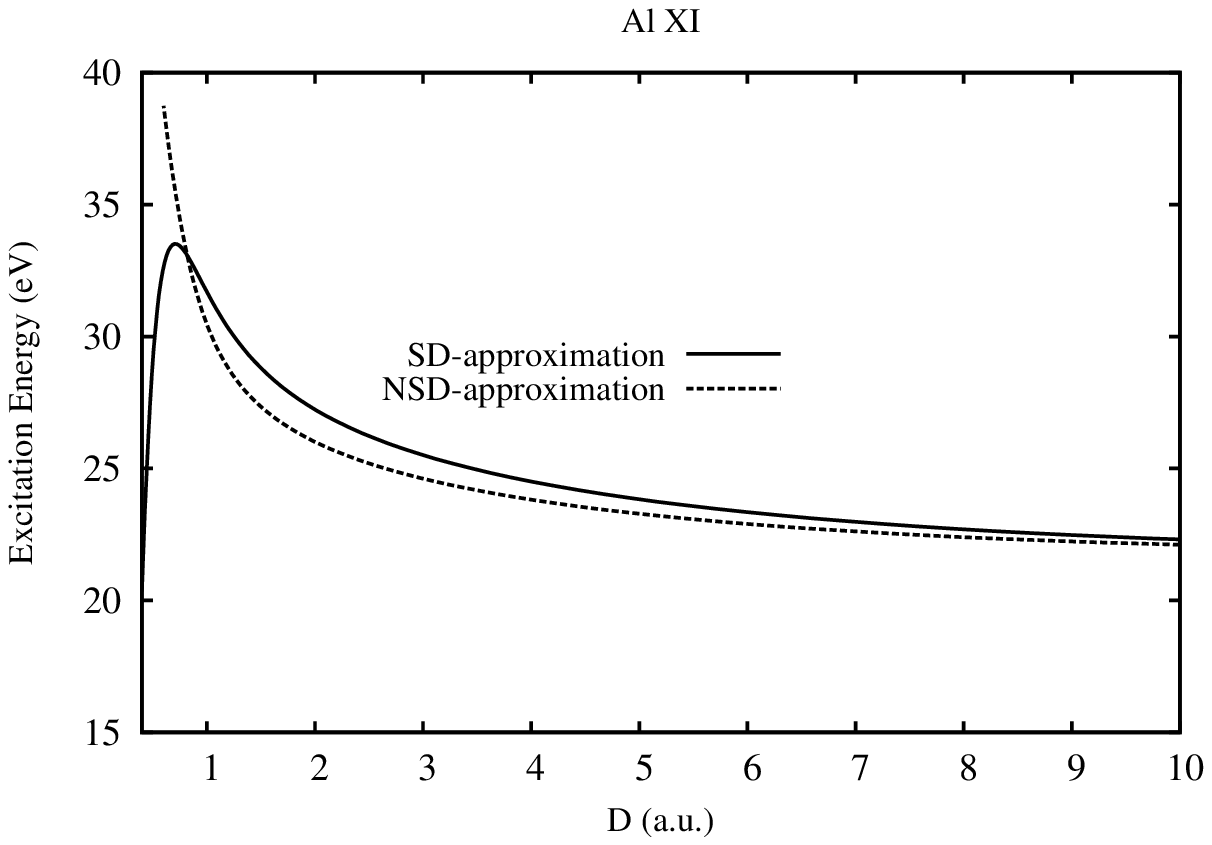}\\
      (iii) & (iv) \\
     \end{tabular}
\caption{Variation in the first EE (in cm$^{-1}$) of (i) Al I, (ii) Al III, (iii) Al IX and (iv) Al XI with the Debye screening length $D$ (in a.u.).}
 \label{fig2}
\end{figure*}

\section{Method of calculations}\label{sec3}

To carry out the atomic wave function calculations in the considered Al-species, we use the Hamiltonian in the SD model given by
\begin{eqnarray}
H &=& \sum_{i=1}^{N} \left [ c\vec{\alpha_{i}}\cdot\vec{p}_{i}+ (\beta-1) c^{2}+ e^{-r_{i}/D}  V_{nuc}(r_{i})\right] \nonumber \\
 & & +  \frac{1}{2} \sum_{i,j} \frac{1}{r_{ij}},
\label{SDeq}
\end{eqnarray}
in the NSD model given by
\begin{eqnarray}
H &=& \sum_{i=1}^{N} \left [ c\vec{\alpha_{i}}\cdot\vec{p}_{i}+ (\beta-1) c^{2}+ e^{-r_{i}/D} V_{nuc}(r_{i})\right] \nonumber \\
  && +  \frac{1}{2} \sum_{i,j} \frac{e^{-r_{ij}/D} }{r_{ij}}  
  \label{NSDeq}
\end{eqnarray}
and 
in the IS model given by 
\begin{eqnarray}
H &=& \sum_{i=1}^{N} \left [ c\vec{\alpha_{i}}\cdot\vec{p}_{i}+ (\beta-1) c^{2}+ V_{\mathrm{eff}}^{\mathrm{IS}}(r_i) \right] \nonumber \\
  && +  \frac{1}{2} \sum_{i,j} \frac{1}{r_{ij}},
\label{ISeq}
\end{eqnarray}
where $\vec{\alpha}$ and $\beta$ are the Dirac matrices and $c$ is the velocity of light. 

The wave functions of the states of the considered atomic systems are evaluated by classifying the orbitals into 
a closed core and a valence orbital. In this approach the wave functions are expressed in the RCC ansatz as 
\begin{eqnarray}
 \vert \Psi_v \rangle &=& e^T \{ 1+ S_v \} \vert \Phi_v \rangle = e^T \{ 1+ S_v \} a_v^{\dagger} \vert \Phi_c \rangle ,
 \label{eqcc}
\end{eqnarray}
where $T$ and $S_v$ are the RCC excitation operators that excite electrons from the core and core along with the valence orbitals to the virtual 
space respectively. Here $\vert \Phi_c \rangle$ and $\vert \Phi_v \rangle$ are the Dirac-Hartree-Fock (DHF) wave functions of the closed-core 
and the closed-core with the valence orbital, respectively. In this work, we have considered only the single and double excitations, denoted by the 
subscripts $1$ and $2$ respectively, in the RCC calculations (known as CCSD method) by expressing
\begin{eqnarray}
 T=T_1 +T_2 \ \ \ \text{and} \ \ \ S_v = S_{1v} + S_{2v}.
\end{eqnarray}
 The amplitudes of these operators are evaluated using the equations
\begin{eqnarray}
 \langle \Phi_c^* \vert \overline{H}_N  \vert \Phi_c \rangle &=& 0
\label{eqt}
 \end{eqnarray}
and 
\begin{eqnarray}
 \langle \Phi_v^* \vert \big ( \overline{H}_N - \Delta E_v \big ) S_v \vert \Phi_v \rangle &=&  - \langle \Phi_v^* \vert \overline{H}_N \vert \Phi_v \rangle ,
\label{eqsv}
 \end{eqnarray}
where $\vert \Phi_c^* \rangle$ and $\vert \Phi_v^* \rangle$ are the excited state configurations, here up to doubles, with respect to the 
$\vert \Phi_c \rangle$ and $\vert \Phi_v \rangle$ DHF wave functions respectively and $\overline{H}_N= \big ( H_N e^T \big )_l$ with subscript 
$l$ represents for the linked terms only. In the above expression, $\Delta E_v$ is the attachment energy of the electron in the valence orbital $v$.
In the {\it ab initio} approach, the $\Delta E_v$ value is determined using the expression
\begin{eqnarray}
 \Delta E_v  = \langle \Phi_v \vert \overline{H}_N \left \{ 1+S_v \right \} \vert \Phi_v \rangle .
 \label{eqeng}
\end{eqnarray}
As can be seen, both Eqs. (\ref{eqsv}) and (\ref{eqeng}) are needed to be solved simultaneously. Hence, Eq. (\ref{eqsv}) is also non-linear 
in $S_v$ operator though it does not appear to be so.

\section{Results and Discussion}\label{sec4}

In this section, we present results obtained for IPs and EEs of Al ions obtained using the RCC method separately for Debye and ion sphere potentials. A detailed 
comparative analysis have been made for the results from both the SD and NSD potential approximations. In order to verify accuracies in our 
results obtained employing the RCC method, we compare the IPs and EEs of the isolated Al atom and its ions with the listed values of National Institute of Science and Technology (NIST) database \cite{NIST} in Table \ref{tab1}. This is done except for 
the EEs of the $3P_{3/2}$ and $4P_{3/2}$ states of Al IX, where the NIST data are not available. We also determine the fine structure splittings (FSs) from these 
values and present them in the same table. We observe good agreement between the calculated and the experimental results except among the 
FS transitions. This may be due to the fact that higher order relativistic correlations are expected to contribute to these transitions substantially.
Agreement among the EEs of the other transitions is an indication that the determined IPs, EEs and FSs of the plasma embedded Al-systems can also be 
of similar accuracies by taking confidence on the validity of the considered models.

\begin{figure}[t]
 \includegraphics[width=8.4cm, height=6.7cm]{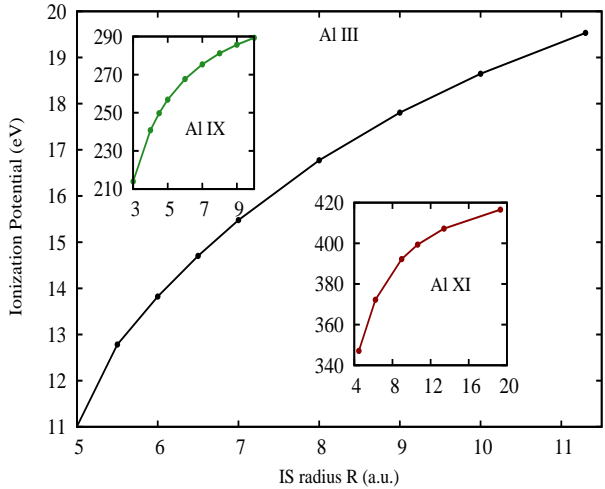}
\caption{Variation of ionization potential (IP) in the Al III ion the with ion sphere radius $R$ (in a.u.). Inset plots are shown for the Al IX and Al XI
ions.}
\label{fig3}
\end{figure}

\subsection{Results from the Debye model}

We perform the calculations of IPs by varying $D$ value, that corresponds to different plasma density ($n_i$) 
and plasma temperature ($T$), in the Debye model. As described,  
the plasma screening effects are included using both the SD potential and NSD potential approximations. A comparison
among the results from these two approximations can demonstrate role of plasma screening through the electron-electron correlation effects
in the considered systems. To examine this, we vary the $D$ value from 5.0 a.u. to 100 a.u. in Al I, from 3 a.u. to 100 a.u. in Al III, 
from 0.6 a.u. to 13.6 a.u. in Al IX, and from 0.5 a.u. to 13.6 a.u. in Al XI. In Fig. \ref{fig1}, we show changes in IPs with the $D$ 
values in Al I, Al III, Al IX and Al XI for both the SD and NSD potential approximations. As expected, the IPs decrease smoothly with 
decrease in the $D$ value in all the systems; this is one of the unique properties of the plasma embedded atomic systems \cite{Inglis,Feynman,Ecker,Stewart}. One can clearly observe from these plots that, the differences in IPs between the SD and NSD potential approximations are 
large in Al I and gradually it gets reduced when Al is more ionized. In Al I, the results from both the approximations are differing substantially
implying it is imperative to include electron correlation effects accurately in the many-electron systems. It can also be noticed from the 
results that the differences in Al I and Al III are slightly larger for the intermediate range of $D$. This may be because of the fact that 
with the increase of screening effect, the electrons become more relaxed at the intermediate values of $D$. When the $D$ value is increased 
further, the electrons start seeing stronger plasma screening effects through the nuclear potential. Hence, for the large $D$ values the 
differences between the IPs gradually decrease in both the SD and NSD potential approximations. To get a quantitative realization of 
variation of IPs with the $D$ values, we quote IPs of all the considered ions for some selective values of $D$ in Table \ref{tab1} using
both the SD and NSD potential approximations. Differences between the results from both the approximations have been given 
as $\Delta$IP in the same table.

\begin{figure}[t]
\includegraphics[width=8.4cm, height=6.7cm]{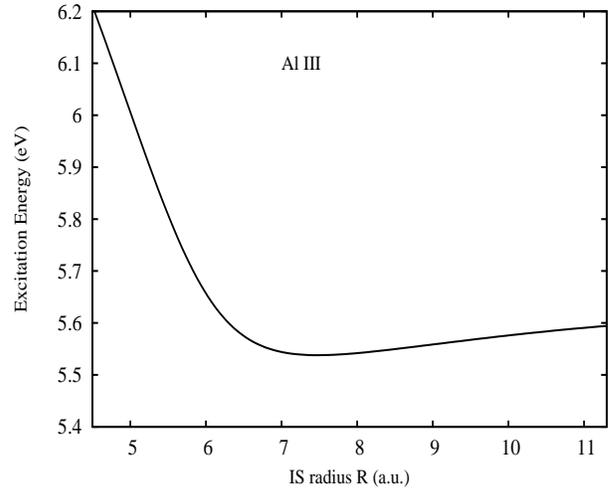}
\caption{Variation of EE of the 3P$_{1/2}$ state in the Al III ion with the IS radius R (a.u.).}
\label{fig4}
\end{figure}

From Fig. \ref{fig1}, we obtain the IPDs to be 2.52 eV, 19.85 eV, 233.41 eV, and 368.95 eV at the $D$ 
values 10 a.u, 3 a.u, 0.8 a.u and 0.5 a.u. for Al I, Al III, Al IX and Al XI, respectively, in the NSD potential approximation. 

\begin{figure}[t]
 \includegraphics[width=8.4cm, height=6.7cm]{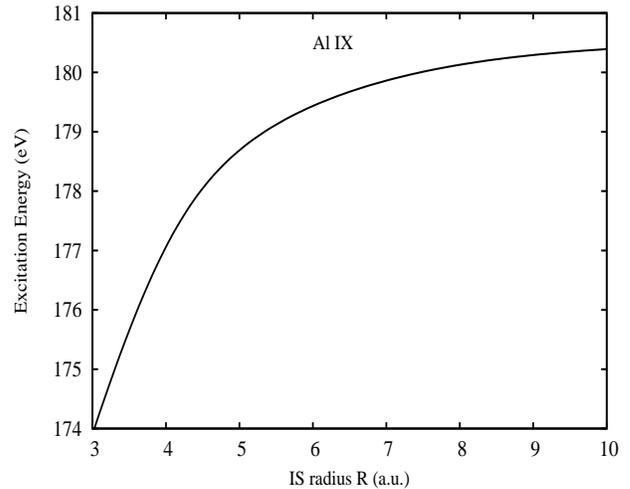}
\caption{Variation of EE of the 3S$_{1/2}$ state in the Al IX ion with the IS radius R (a.u.). }
\label{fig5}
\end{figure}

To understand the variation in the excitation energies of plasma embedded ions, we also investigate variation in the first excited state energies 
of the Al I, Al III, Al IX and Al XI ions, with the Debye length. In Fig. \ref{fig2}, we have plotted them against the $D$ values considering both the 
SD and NSD potential approximations. One can infer from this figure that the EEs of the excited states in the plasma embedded Al systems decrease 
with decreasing values of $D$ except in the Al XI ion. We find the situation is quite different in the Al XI ion, where the EE of the $2P_{1/2}$ state 
increases with the increasing screening strength. However at very high screening region, this EE starts decreasing at some critical $D$ value in 
the SD approximation. This behavior was also seen in the hydrogen-like ions \cite{Qi2008} and lithium-like ions \cite{mdas2014} in the Debye model 
studies. Where as in the NSD potential approximation, this behavior disappears. Therefore, it implies that the use of NSD potential in the Debye model 
reduces the electron-electron screening effects that are overestimated in the SD model approximation in the systems having many electrons. Thus, it demonstrates importance of accounting the plasma screening effects through the two-body 
interaction term in the many-electron systems accurately. Another aspect can be observed from the analysis of EEs in the considered systems is that
the differences between the EEs obtained using the SD and NSD potential approximations are larger than their IPs, as seen in Figs. \ref{fig1} 
and \ref{fig2}. This means that the screening effects affect the ground and the excited states differently.

\subsection{Results from IS model}

Here, we proceed presenting results from the IS model by carrying out the calculations using the RCC method. These results are supposed to explain 
the systems in the strong coupling plasma environment. In Fig. \ref{fig3}, we have plotted the IPs of the Al III, Al IX and Al XI ions with 
different IS radii. In this model, the IS radius is varied from 4 a.u. to 11.3 a.u., 3 a.u. to 10 a.u. and 4.48 a.u. to 19.32 a.u. in the  
Al III, Al IX and Al XI ions, respectively. From the plot we can see that the IPs decrease monotonically with decreasing the ion 
sphere radius $R$. The decaying trends in the plots mean growing instability in the system with the rise of ion density in the strong plasma 
environment. Similar pattern was also observed by Sil {\it et al} \cite{Sil} for the plasma embedded Al$^{11+}$ ion in the same IS model analysis.

\begin{figure}[t]
 \includegraphics[width=8.4cm, height=6.7cm]{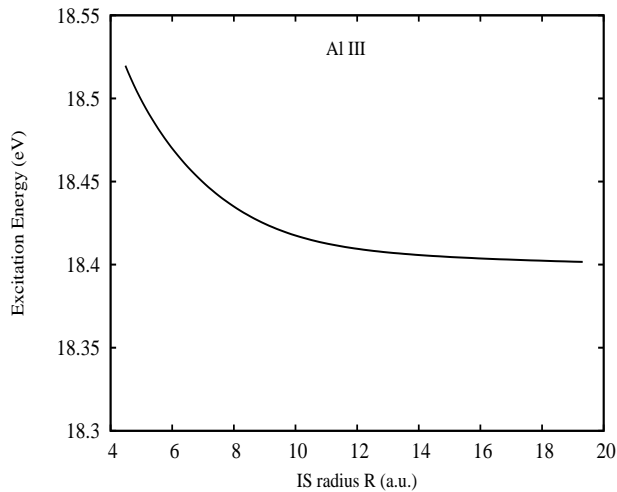}
\caption{Variation of EE of the 2P$_{1/2}$ state in the Al XI ion with IS radius R (a.u.).}
\label{fig6}
\end{figure}

In the same spirit as of Debye model, we have also investigated the trends in the EEs of the first excited states of all the considered
atomic systems in the IS model by varying $R$. The variation of the EE of the $3P_{1/2}$ state in the Al III ion with IS radius shows very 
peculiar result as shown in Fig. \ref{fig4}. Initially, the EEs of the $3P$ states decrease with decreasing value of IS radius, then they suddenly 
rise at certain critical values of $R$. Similar trends are also seen in the higher excited states of the Al III ion.

In Figs. \ref{fig5} and \ref{fig6}, we have plotted the variation of EEs in the Al IX and Al XI ions, respectively. From Fig. \ref{fig5}, we 
find that in the Al IX ion EE of the $3S_{1/2}$ state decreases with decreasing value of IS radius. Whereas in the Al XI ion, EE of the $2P_{1/2}$ 
state increases with decreasing radius of IS. From these findings in the high density plasma, we conclude that for low ionized systems, like 
Al III, the transition spectra are initially red shifted and towards very high density region, it becomes blue shifted. In contrast, the highly 
ionized ions, such as Al IX and Al XI, show blue shift in the spectral lines in the transitions between the states having same principal quantum 
numbers and red shifted in the transitions involving states with different principal quantum numbers.

\begin{table}[t]
\caption{Comparison of EEs (Case I) obtained using the IS model with radius $R=20$ a.u. and Debye model with NSD potential approximation for $D=11.3$ a.u. that corresponds to 
the experimental conditions with $T=20$ eV and $n_i=0.2 \times 10^{21}$/cc. In another case (Case II), the above quantities are compared using the IS model with radius 
$R=16$ a.u. and Debye model with NSD potential approximation for $D=13.6$ a.u. corresponding to the experimental conditions with $T=(58 \pm 4)$ eV and $n_i=0.4 \times 10^{21}$/cc.}\label{tab3}
\begin{ruledtabular}
\begin{tabular}{c c c c c }
    &              &   Experimental            &\multicolumn{2}{c}{ EE (in cm$^{-1}$)} \\
    \cline{4-5} \\
Ion & State       & condition &  IS model  & Debye model  \\
\hline
 & & \\
Al III & 3P$_{1/2}$ &          & 45349.90   &52819.41\\
       & 3P$_{3/2}$ &   & 45534.82   &53056.49\\
       & 3D$_{3/2}$ & Case I & 100971.21  &113316.78\\
       & 3D$_{5/2}$ &  & 100970.98  & 113315.68\\
       & 4S$_{1/2}$ &     & 115413.04  &120263.59\\
       & 4P$_{1/2}$ &         & 130384.45   &136394.76\\
       & 4P$_{3/2}$ &         & 130447.22   &136468.07\\
      \hline
      & & \\
Al XI  &  3S$_{1/2}$&         &  1995874.47 &2016739.20\\
       &  3P$_{1/2}$&         &  2036363.33 &2065474.87\\
       &  3P$_{3/2}$& Case II &  2038140.30 &2067306.84\\
       &  3D$_{3/2}$&   &             &2085870.23\\
       &  3D$_{5/2}$&     &             &2086399.55\\
       &  4S$_{1/2}$&         &  2679621.16 &2697199.34\\
       &  4P$_{1/2}$&         &  2692124.54&2717306.72\\
       &  4P$_{3/2}$&         &  2692875.54&2718069.23\\
\end{tabular} 
\end{ruledtabular}
\end{table}

\subsection{Debye versus IS model results}

Though it is well known that Debye model describes well the weakly coupled plasma and IS model describes appropriately the strongly coupled plasma, 
we just wanted to investigate validity of both the models in the similar plasma conditions. For this purpose, we have calculated the IPDs of the
Al III and Al XI ions under the same experimental plasma conditions using both the Debye model and IS model at the intermediate plasma coupling. 
Under the experimental condition 
\cite{Winhart} with the ion density $n_i = 0.2 \times 10^{21}/cc$ and the temperature 20 eV (which correspond to $D=11.36$ a.u. and $R=20$ a.u.), 
the obtained IPDs of the Al III ion are 6.6 eV and 5.8 eV in the Debye and IS models, respectively. Similarly, with the experimental plasma condition 
\cite{Perry} with $n_i = 0.4 \times 10^{21}/cc$ and $T = 58 \pm 4 $ eV (corresponding to $D=13.6$ a.u. and $R=16$ a.u.), the IPD of Al XI are
found to be 21.60 eV and 24.3 eV in the Debye and IS models, respectively. We have also estimated EEs of many low-lying excited states of the 
Al III and Al XI ions using these two models under the above experimental conditions and given them in Table \ref{tab3} for the comparison purpose. 
We find a quite good agreement in the results from both the models in Al III and Al XI, while results from Debye model are relatively larger.
The above plasma coupling strengths under the experimental conditions ($n_i = 0.2 \times 10^{21}/cc, T=20eV$) and ($n_i = 0.4 \times 10^{21}/cc$, 
$T = 58 \pm 4  eV$) are about 2.8, and 1.2, receptively, they are in the intermediate range of classifying as either weakly coupled plasma 
(i.e. $\Gamma$ $\ll$ 1) or strongly coupled plasma (i.e. $\Gamma$ $\ge$ 10). So, one would expect in this intermediate region, both the models need 
to give comparatively the similar results. We anticipate that results obtained using the IS model are more valid here as the plasma couplings 
corresponding to the above plasma conditions are larger than 1, where Debye model may not be able to describe the systems appropriately.  

\section{Concluding remarks}\label{sec5}

In this work, we have investigated the electronic structures of Al atom and some of its ions both in the weakly and strongly coupling plasma 
environments considering the Debye and ion-sphere models, respectively. Further, we have investigated differences in the results considering the 
plasma screening effects in the electron-electron repulsion with the spherical potential and non-spherical potential approximations within the Debye
model to estimate ionization potential depressions and excitation energies of the considered systems. We find significant differences in the results 
in the systems having more electrons. It also predicts more stability in the atomic systems when the exact screening effects are taken into
account.  A similar analysis has also been carried out to analyze structures of the Al ions in the strongly coupled plasma using the 
ion sphere model. In this model, we find the highly ionized ions show blue shifts in the transitions among the states with same principal
quantum numbers, else red shifts with the decreasing values of ion sphere radius. Out of keenness, we have also applied both the Debye and ion-sphere
models to carry out calculations of the ionization potential depressions and excitation energies in the Al III and Al XI ions considering the 
plasma conditions that fall in the intermediate coupling plasma strengths. We find from this investigation that the ionization potential depressions 
for the Al III and Al XI ions obtained using both the plasma models reasonably agree each other, but Debye model predicts higher values. 

\section*{Acknowledgment}

 M. Das thanks CSIR for financial support. A part of the computations were carried out using Vikram-100 HPC cluster at Physical Research Laboratory, 
Ahmedabad.

\end{document}